\documentclass[11pt]{article}
\usepackage[utf8]{inputenc}
\usepackage[T1]{fontenc}
\usepackage{amsmath,amssymb,graphicx}
\usepackage{geometry}
\usepackage{setspace}
\usepackage{caption}
\usepackage{indentfirst}

\usepackage{caption}
\usepackage{url}
\usepackage{xurl} 

\usepackage{xcolor}
\usepackage{hyperref}

\hypersetup{
    colorlinks=true,
    linkcolor=black,
    citecolor=black,
    urlcolor=blue
}

\begin{document}

\title{Two-branch retention behavior in unsaturated fractured rock driven by fracture--matrix flow partitioning}

\author{
\begin{tabular}{c}
\large M.R. Andiva$^{1}$, C. Jiang$^{1}$, M. Ziegler$^{2}$, Q. Lei$^{1}$ \\[0.6em]
\normalsize $^{1}$Department of Earth Sciences, Uppsala University, Uppsala, Sweden \\
\normalsize $^{2}$Swiss Federal Office of Topography (swisstopo), Mont Terri URL, St-Ursanne, Switzerland \\[0.5em]
\small *Correspondence: Muhammad Raharsya Andiva 
(\href{mailto:muhammad.andiva@geo.uu.se}{muhammad.andiva@geo.uu.se})
\end{tabular}
}

\date{}
\maketitle

\section*{Abstract}
Upscaling unsaturated flow in fractured rock remains challenging because fractures and matrix often exhibit sharply contrasting hydraulic behaviors across saturation states. Here, we demonstrate that unsaturated flow undergoes a transition between matrix- and fracture-dominated regimes. Three-dimensional direct numerical simulations reveal that both relative permeability and capillary pressure curves display a robust two-branch structure. We analytically derive a generalized retention formulation that identifies a critical saturation marking the transition between the two distinct retention regimes and reproduces the two-branch behavior captured in the numerical simulations. An analytical expression for the critical pressure head is further derived to represent the limiting case of fully connected fracture networks, providing a physical explanation for the retention regime shift and showing good agreement with the numerical results for systems above the percolation threshold. Our results provide a mechanistic framework for understanding and upscaling unsaturated flow in fractured rock, with broad implications for hydrology and geophysics.

\section{Introduction}
Unsaturated flow in fractured rock plays a critical role in subsurface fluid transport, with important implications for groundwater management and nuclear waste disposal (Pruess \& Tsang, 1990; Liu et al., 1998; Nimmo, 2012; Rutqvist \& Tsang, 2012; Zhang, 2018; Wang et al., 2019; Liang et al., 2022; Viswanathan et al., 2022). However, predicting unsaturated flow in such systems remains challenging, because fractured media consist of two distinct domains, i.e., the rock matrix and the fracture network, with strongly contrasting hydraulic and retention properties (Glass et al., 1995; Seol et al., 2003; Huang et al., 2005). Unlike saturated conditions, where fractures typically dominate fluid flow, under unsaturated conditions, the relative importance of matrix versus fracture flow remains unresolved.

From a theoretical perspective, the porous rock matrix is expected to play a dominant role under unsaturated conditions, because strong capillary forces promote imbibition into the fine pore space of the matrix, which can suppress rapid flow through fractures and lead to relatively diffuse infiltration (Wang \& Narasimhan, 1985; Nitao \& Buscheck, 1991). Based on this conceptual view, subsurface flow has been traditionally represented using continuum approaches that treat fractured rock as an equivalent porous medium (Peters \& Klavetter, 1988). However, field observations have revealed a different behavior. Field experiments at sites such as Yucca Mountain (Nevada, United States) have shown that water can migrate rapidly through localized preferential pathways along fractures even under unsaturated conditions (Pruess, 1998; Pruess et al., 1999; Bodvarsson et al., 2003). Laboratory experiments have also illustrated that flow can proceed with minimal imbibition into the surrounding rock matrix, thereby limiting fracture--matrix hydraulic exchange and enabling fast fracture-dominated transport (Liu et al., 1998; Wood et al., 2002; Ji et al., 2004; Tokunaga et al., 2005; Zhou et al., 2022). Furthermore, the geometry and aperture distribution of fracture networks strongly influence recharge efficiency and the rate at which water percolates through the unsaturated zone (Birkholzer et al., 1999; Liu et al., 2002; Cey et al., 2006; Sherman et al., 2020; Zhu et al., 2021).

These concepts and findings expose a paradox: capillary forces favor matrix-dominated flow, yet fractures serve as preferential pathways for unsaturated flow. Here, we address this paradox by showing that flow in fractured rock exhibits a two-branch retention behavior, reflecting a transition between matrix-dominated and fracture-dominated regimes. Combining three-dimensional numerical simulations with analytical development, we derive a generalized retention formulation that captures this transition behavior and reveals the underlying physical mechanisms. These results provide a more complete and physically based picture of unsaturated flow in fractured rock.

\section{Methodology}
Unsaturated flow through fractured porous media can be described by the Richard's equation (Bear, 1972). Under steady-state conditions, neglecting gravity and in the absence of sinks or sources, the governing equations for the matrix and fracture domains are respectively written as:

\begin{equation}
\nabla \cdot \left(-\rho \frac{k_m k_{rm}}{\mu} \nabla p \right) = 0
\tag{1}
\end{equation}

\begin{equation}
\nabla_{\tau} \cdot \left(-b\rho \frac{k_f k_{rf}}{\mu} \nabla_{\tau} p \right) = \rho \left(q^{+} + q^{-}\right)
\tag{2}
\end{equation}

\noindent where $p$ is fluid pressure, $\rho$ is fluid density, $\mu$ is dynamic viscosity, $b$ is fracture aperture, $k$ is absolute permeability, $\nabla_{\tau}$ denotes the tangential gradient operator along the fracture plane, and $k_r$ is relative permeability, with the subscripts ``m'' and ``f'' indicating matrix and fracture properties, respectively. Fracture permeability $k_f$ is calculated based on the cubic law (Witherspoon et al., 1980), i.e., $k_f = b^2/12$. The term $\rho(q^{+} + q^{-})$ in Equation 2 represents fracture-matrix mass exchange, with the interfacial fluxes defined as $q^{+} = -(k_m k_{rm}/\mu)(\partial p^{+}/\partial n^{+})$ and $q^{-} = -(k_m k_{rm}/\mu)(\partial p^{-}/\partial n^{-})$, where $n^{+}$ and $n^{-}$ denote the outward normal directions on the respective fracture faces.

Unsaturated flow behavior is additionally controlled by constitutive relationships linking capillary pressure $p_c$, effective saturation $S_e$, and relative permeability $k_r$. The $p_c$--$S_e$ relationship is commonly referred to as the retention curve. Here, we adopt the Brooks-Corey model (Brooks \& Corey, 1966) with separate parameter sets specified for the fracture and matrix domains. Expressed in terms of pressure head $h$, the Brooks--Corey formulation is written as:

\begin{equation}
S_e =
\begin{cases}
\dfrac{1}{\lvert \alpha h \rvert^n}, & h < -\dfrac{1}{\alpha} \\[0.5em]
1, & h \ge -\dfrac{1}{\alpha}
\end{cases}
\tag{3}
\end{equation}

\begin{equation}
k_r =
\begin{cases}
S_e^{\frac{2}{n}+l+2}, & h < -\dfrac{1}{\alpha} \\[0.5em]
1, & h \ge -\dfrac{1}{\alpha}
\end{cases}
\tag{4}
\end{equation}

\begin{equation}
p_c = \frac{\rho g}{\alpha} S_e^{-\frac{1}{n}}
\tag{5}
\end{equation}

\noindent where $g$ is the gravitational acceleration, $\alpha$ is the inverse of the air entry pressure head, and $n$ and $l$ are constants. While several alternative retention models (van Genuchten, 1980; Rogers et al., 1983) exist, we adopt the Brooks-Corey formulation because it is well suited to rock materials (Assouline, 2005; Gershenzon et al., 2016). To demonstrate the generality of our framework, we additionally test the van Genuchten formulation, with results presented in Text S5 and Figure S12 in Supporting Information.

We construct a series of three-dimensional synthetic fracture networks within a cubic domain of side length $L$, where fracture radii $r$ follow a power-law distribution (Bonnet et al., 2001), with the density function given by:

\begin{equation}
f(r,L) = \xi r^{-a}, \quad r_{\min} \le r < r_{\max}
\tag{6}
\end{equation}

\noindent where $a$ is the power-law exponent, $\xi$ is a density term, and $r_{\min}$ and $r_{\max}$ are the minimum and maximum fracture radii. Fracture orientations and locations are assumed to be purely random, representing nominally isotropic conditions. Fracture networks are characterized by the fracture intensity $\gamma$ (defined as total fracture area per unit volume) and the percolation parameter $\chi$ (defined as normalized excluded volume), which are respectively related to the second and third moments of the density function as follows (Bour \& Davy, 1998; de Dreuzy et al., 2000):

\begin{equation}
\gamma(r',L) = \int \frac{f(r,L)\pi r'^2}{L^3} \, dr'
\tag{7}
\end{equation}

\begin{equation}
\chi(r',L) = \int \frac{f(r,L)\pi^2 r'^3}{L^3} \, dr'
\tag{8}
\end{equation}

\noindent where $r'$ is the effective fracture radius corresponding to the portion of a fracture contained within the finite domain. The percolation parameter $\chi$ is a metric quantifying fracture network connectivity, where the system transitions from disconnected to connected states as $\chi$ goes beyond the percolation threshold $\chi_c$ having a value is around 0.7 to 2.8 (Balberg, 1985; de Dreuzy et al., 2000).

We consider five power-law exponents ($a = 2.5$, 3, 3.5, 4, and 4.5) and five fracture intensities ($\gamma =2.5/L$, $5/L$, $10/L$, $20/L$, and $40/L$), resulting in 25 parametric combinations (Figure 1). Ten realizations are generated for each case. Fractures are assigned with a constant aperture of $L/1000$ and a porosity of unity (i.e., no infilling material). The surrounding matrix has a permeability of $10^{-18} \times L^2$ and a porosity of 0.01. Distinct Brooks-Corey parameters are assigned to fractures and matrix. For fractures, the inverse air-entry suction parameter $\alpha_f$ is related to fracture aperture, contact angle, and fluid properties through Jurin's law (Ababou, 2018), expressed as follows:

\begin{equation}
\alpha_f = \frac{b}{2\sigma_w \cos \theta} {\rho g}
\tag{9}
\end{equation}

\noindent where $\sigma_w$ is the water-air surface tension and $\theta$ is the contact angle. The matrix pore radius is significantly smaller than the fracture aperture with $\alpha$ for matrix set to be $1/L$. The Brooks-Corey parameter $n$ was assigned to be 2 for the matrix and 1.5 for the fractures, consistent with reported lower values for fractures relative to the matrix (Li et al., 2014).

A series of steady-state simulations was conducted to explore different saturation states. To span the full range of effective saturation, the reference pressure head imposed at the inflow boundary is systematically varied using 126 logarithmically spaced values, ranging from $-0.001L$ to $-2000L$, with locally refined spacing over a selected interval to improve resolution near the transition. A uniform hydraulic gradient of $-0.001$ is imposed in the flow direction along the x-axis. The remaining four lateral boundaries were treated as impermeable. For each simulation, the resulting capillary pressure and relative permeability fields are extracted for subsequent analysis. Air pressure is assumed constant, such that negative water pressure corresponds directly to capillary pressure. The upscaled relative permeability is determined from the total volumetric outflow rate $Q$ at the outlet boundary, accounting for contributions from both fracture and matrix domains. The absolute permeability $k_{\mathrm{abs}}$ is obtained from a fully saturated simulation. By rearranging Equation 1, the relative permeability of the system can be obtained as:

\begin{equation}
k_r = - \frac{Q\mu}{k_{\mathrm{abs}} L \Delta P}
\tag{10}
\end{equation}

\noindent where $\Delta P$ is the pressure differential imposed across the system. All results and analyses in our study are presented in dimensionless form to ensure generality.

\begin{figure}[htbp]
\centering
\includegraphics[width=\textwidth]{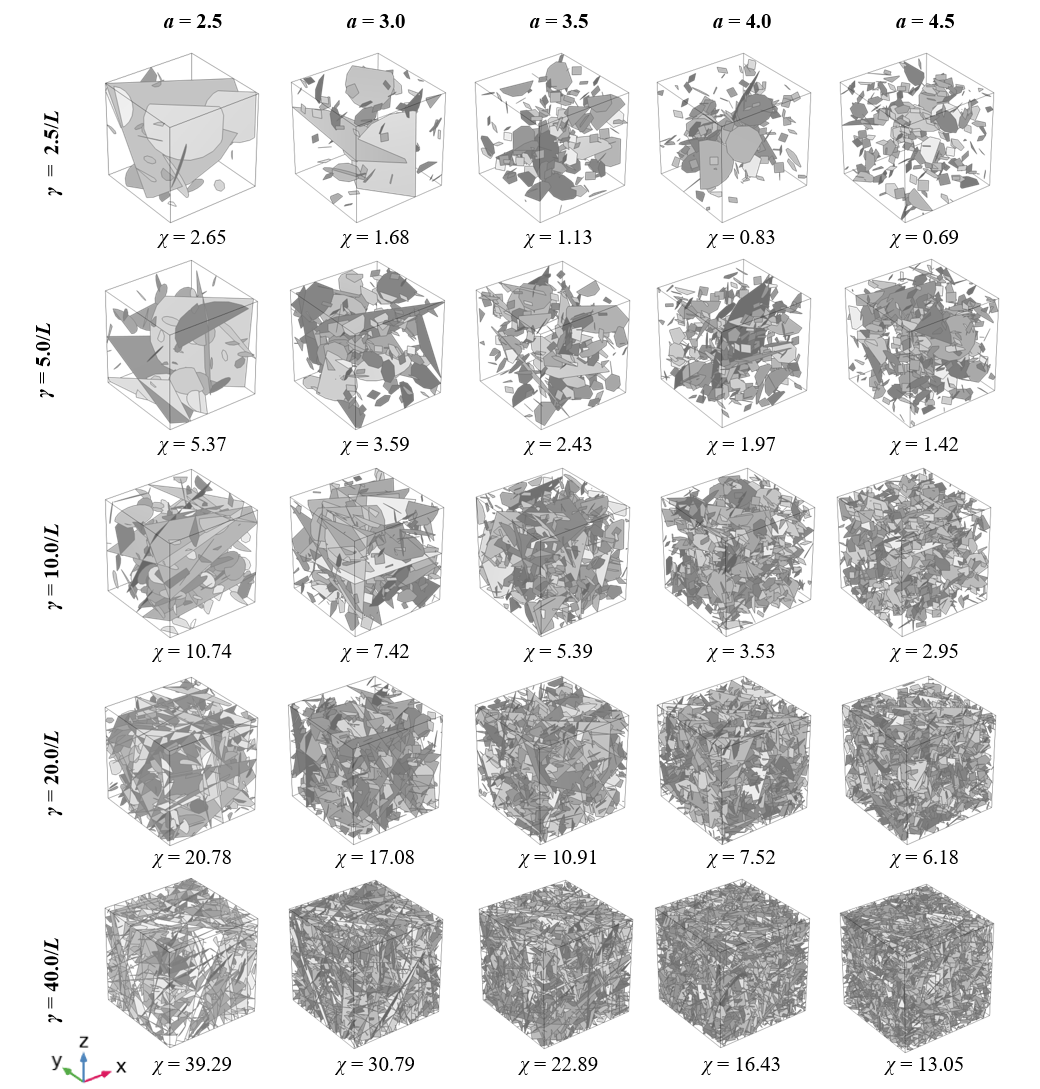}
\caption{Generated three-dimensional fracture networks associated with different power-law exponents of fracture radii $a$ and fracture intensities $\gamma$. Ten realizations are generated for each case, with only one representative realization presented here for illustration. The percolation parameter $\chi$ is also computed to quantify fracture network connectivity.}
\end{figure}

\section{Results}
\subsection{Emergent two-branch retention behavior}

Figure 2 shows the upscaled relative permeability and capillary pressure curves for a representative fracture network case ($a = 3$, $\gamma = 20/L$; Figure 1), with the contributions from the fracture and matrix domains also shown. Both the relative permeability (Figure 2a) and normalized capillary pressure $p_c/\rho gL$ (Figure 2b) display a distinct two-branch structure, arising from the contrasting hydraulic behavior of the fracture and matrix domains across the saturation range. At low effective saturations, flow is dominated by matrix. As saturation increases, the system transitions to a regime dominated by fracture flow. The fracture contribution sharply increases at $S_e \approx 0.37$, referred to hereafter as the critical saturation $S_c$, which marks the transition from matrix-dominated to fracture-dominated regimes. This two-branch retention behavior is consistently observed across the broad range of fracture network scenarios examined in this study, including all stochastic realizations generated (Text S3 and Figures S8 and S9 in Supporting Information).

\begin{figure}[htbp]
\centering
\includegraphics[width=\textwidth]{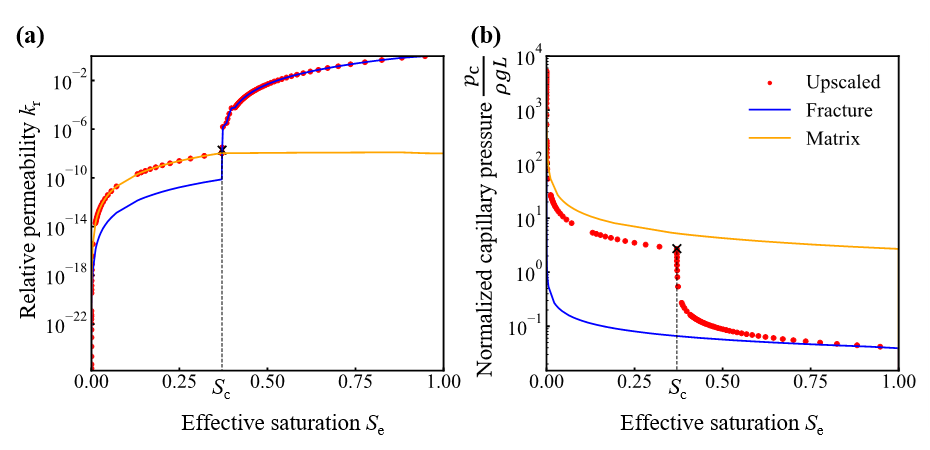}
\caption{Numerical simulation results of upscaled (a) relative permeability $k_r$ and (b) normalized capillary pressure $p_c/\rho gL$ as functions of effective saturation $S_e$ in a fractured rock with the power-law exponent $a = 3$ and fracture intensity $\gamma = 20/L$. Matrix and fracture contributions are also separately shown to illustrate their relative roles. The black cross indicates the critical saturation $S_c$, marking the transition between matrix-dominated and fracture-dominated flow regimes.}
\end{figure}

We propose a two-branch Brooks-Corey model that captures the transition between matrix- and fracture-dominated flow regimes, given by:

\begin{equation}
k_r(S_e) =
\begin{cases}
k_{rc}\left(\dfrac{S_e}{S_c}\right)^{\frac{2}{n_m}+l_m+2}, & S_e \le S_c \\[0.75em]
k_{rc} + (1-k_{rc})\left(\dfrac{S_e-S_c}{1-S_c}\right)^{\frac{2}{n_f}+l_f+2}, & S_e > S_c
\end{cases}
\tag{11}
\end{equation}

\begin{equation}
p_c(S_e) =
\begin{cases}
p_{cc}\left(\dfrac{S_e}{S_c}\right)^{-1/n_m}, & S_e \le S_c \\[0.75em]
p_{cc}\left(\dfrac{S_e-S_c}{1-S_c}+\varepsilon\right)^{-1/n_f}\varepsilon^{1/n_f}, & S_e > S_c
\end{cases}
\tag{12}
\end{equation}

\noindent where $k_{rc}$ is the relative permeability at $S_c$ and $p_{cc}$ is the capillary pressure at $S_c$. The parameter $\varepsilon$ is introduced to ensure continuity and to regularize the singular behavior at $S_e = S_c$, defined as:

\begin{equation}
\varepsilon = \frac{1}{\left(\dfrac{\alpha_f p_{cc}}{\rho g}\right)^{n_f} - 1}
\tag{13}
\end{equation}

This generalized formulation extends the classical Brooks-Corey relationship by introducing regime-dependent normalization and incorporating the critical saturation (associated with $k_{rc}$ and $p_{cc}$) separating the two regimes. If the effective saturation $S_e$ is below the critical saturation $S_c$, the system resides in the matrix-dominated regime, in which the constitutive relationships are expressed in terms of the normalized saturation $S_e/S_c$. For $S_e > S_c$, corresponding to the fracture-dominated regime, where the constitutive relationships are expressed in terms of the normalized critical saturation $(S_e-S_c)/(1-S_c)$. The constitutive parameters $n$ and $l$ are assigned with matrix properties for $S_e$ below $S_c$, and assigned with fracture properties for $S_e$ above $S_c$. In the limits $S_c \to 0$ and $S_c \to 1$, the generalized formulation reduces to the classical Brooks--Corey model for fracture-only and matrix-only flow, respectively. This two-branch analytical formulation closely matches the numerical simulation results and well captures the regime transition at $S_c$ (Figure 3).

\begin{figure}[htbp]
\centering
\includegraphics[width=\textwidth]{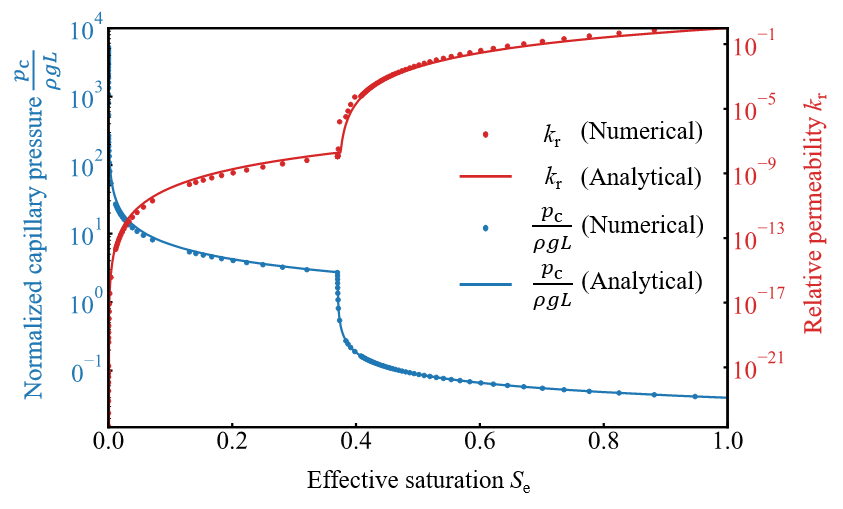}
\caption{Comparison of numerically simulated and analytically predicted relative permeability and capillary pressure curves. The scatter points represent simulation results of relative permeability $k_r$ and normalized capillary pressure $p_c/\rho gL$ as functions of effective saturation $S_e$, while the solid lines indicate the prediction by the two-branch analytical formulation.}
\end{figure}

\subsection{Transition between matrix- and fracture-dominated regimes}

As saturation increases, the matrix and fracture domains evolve differently due to the contrast in their air-entry pressures. The matrix reaches full saturation earlier and attains its maximum flow capacity with the relative permeability $k_{rm}$ reaching unity, while fractures remain partially saturated and their relative permeability $k_{rf}$ continues to increase. Once the matrix becomes fully saturated, the transition to fracture-dominated flow occurs at the point where fracture and matrix fluxes are equal, beyond which fractures dominate the flow.

By combining Equations 3 and 4, the relative permeability $k_r$ can be expressed as:

\begin{equation}
k_r = \left(\frac{1}{|\alpha h|^n}\right)^{\frac{2}{n}+l+2}
\tag{14}
\end{equation}

\noindent Substituting Equation 14 into Darcy’s law yields the volumetric flow rate as a function of pressure head $h$:

\begin{equation}
Q = -\frac{kA}{\mu}\left(\frac{1}{|\alpha h|^n}\right)^{\frac{2}{n}+l+2}\frac{\Delta P}{L}
\tag{15}
\end{equation}

\noindent Using Equation 15 and the condition $\int Q_m dL = \int Q_f dL$, the critical pressure head $h_c$ at this transition is derived as:

\begin{equation}
h_c =
\left[
\frac{k_f \int A_f \, d}{k_m \int A_m \, d}
\frac{\alpha_m^{2+n_m(l_m+2)}}{\alpha_f^{2+n_f(l_f+2)}}
\right]^{\frac{1}{n_f(l_f+2)-n_m(l_m+2)}}
\tag{16}
\end{equation}

\noindent where $A_f$ and $A_m$ denote the cross-sectional areas of the fracture and matrix domains, respectively, normal to the flow direction.

The total fracture cross-sectional area scales with the fracture intensity $\gamma$ and the aperture $b$, whereas the matrix occupies the remainder. Incorporating this geometric partitioning into Equation 16 yields the following analytical expression for the critical pressure head:

\begin{equation}
h_c =
\frac{1}{\alpha_f}
\left[
\frac{k_f b \gamma}{k_m(1-b\gamma)}
\right]^{\eta}
\approx
\frac{1}{\alpha_f}
\left(
\frac{k_f b \gamma}{k_m}
\right)^{\eta}
\tag{17}
\end{equation}

\noindent where

\begin{equation}
\eta = \frac{1}{2+n_f l_f + 2n_f}
\tag{18}
\end{equation}

\noindent Here, we use the approximation $(1-b\gamma)\approx 1$ given that $b\gamma \ll 1$. Combining Equations 9 and 17, and invoking the cubic law for fracture permeability, the critical pressure head $h_c$ can be expressed explicitly as a function of fracture aperture:

\begin{equation}
h_c =
\frac{b^{3\eta-1}2\sigma_w \cos\theta}{\rho g}
\left[
\frac{\gamma}{k_m(1-b\gamma)}
\right]^{\eta}
\approx
\frac{b^{3\eta-1}2\sigma_w \cos\theta}{\rho g}
\left(
\frac{\gamma}{k_m}
\right)^{\eta}
\tag{19}
\end{equation}

\noindent The critical saturation $S_c$ can be further derived as (see Text S1 in Supporting Information for detailed derivations):

\begin{equation}
S_c =
\frac{
\left|
\frac{\alpha_m 2\sigma_w \cos\theta}{b\rho g}
\left(
\frac{b^3\gamma}{12k_m(1-b\gamma)}
\right)^{-\eta n_m}
\right|(1-\gamma b)\phi_m
+
\left|
\left(
\frac{b^3\gamma}{12k_m(1-b\gamma)}
\right)^{-\eta n_f}
\right|\gamma b
}{
(1-\gamma b)\phi_m + \gamma b
}
\tag{20}
\end{equation}

\noindent where $\phi_m$ denotes the matrix porosity.

Figure 4 compares numerical results with the analytical solution (Equation 17) for the critical pressure head $h_c$. Here, we have rearranged Equation 17 to define a dimensionless critical pressure head $h_c \alpha_f/(b\gamma)^{\eta}$, which is plotted as a function of the percolation parameter $\chi$. For $\chi$ above the percolation threshold $\chi_c \approx 0.7-2.8$, the numerical results asymptotically approach the analytical solution, consistent with the assumption of fully connected fractures underlying Equation 17. Around and below the percolation threshold, systematic deviations emerge, because fracture connectivity becomes insufficient to sustain continuous flow pathways. Near the percolation threshold, stochastic variability in network connectivity also leads to coexistence of connected and disconnected realizations, explaining the scatter observed in the numerical results.

\begin{figure}[htbp]
\centering
\includegraphics[width=\textwidth]{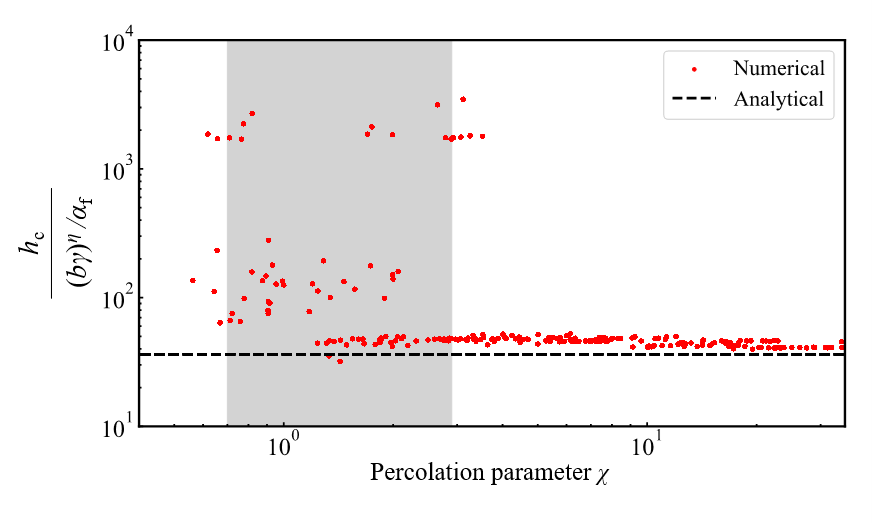}
\caption{Numerical results (red dots) and analytical solution (black dashed line; Equation 16) for the dimensionless critical pressure head plotted against the percolation parameter $\chi$. The shaded gray area indicates the range of the percolation threshold $\chi_c \approx 0.7-2.8$.}
\end{figure}

\section{Discussion}
\subsection{Mechanisms of retention regime transition in unsaturated fractured rock}

The two-branch retention behavior revealed in this study reconciles the longstanding debate over the relative roles of matrix and fracture flow in unsaturated fractured rock. Our results show that both matrix and fractures play important roles, with their relative contributions strongly controlled by saturation. At low saturation, flow rates are small and capillary forces dominate; capillary imbibition draws water into the rock matrix, suppressing fracture flow and regulating air-water distribution (Demond \& Roberts, 1991). As saturation increases, flow rates increase and viscous forces progressively outweigh capillary entry resistance (Xu et al., 1998); the matrix becomes fully saturated first with its relative permeability approaching unity, while fractures---owing to their weaker capillary effects---have not yet reached their maximum flow capacity, i.e., the relative permeability of fractures remains below unity (see Text S4 and Figures S10 and S11 in Supporting Information). As capillary pressure continues to decrease, the contribution of fracture flow to the total flow increases until it becomes comparable to, and eventually greater than, matrix flow. This competition between matrix and fracture flow gives rise to the two-branch retention behavior observed in our study, with the transition between the two regimes defined by a critical saturation $S_c$ at which fracture flow begins to exceed matrix flow. Despite previous experimental studies reporting distinct matrix- and fracture-dominated flow regimes (Nicholl et al., 1994; Tidwell et al., 1995; Liu \& Bodvarsson, 2001; Glass et al., 2003; LaViolette et al., 2003; Rüdiger et al., 2024; Xiang et al., 2025), we here provide a quantitative framework that rationalizes these earlier observations and resolves the apparent paradox regarding the relative importance of matrix and fracture flow.

The retention regime transition discussed above occurs at a critical saturation $S_c$ associated with a critical pressure head $h_c$. Equation 17 shows that this critical saturation is primarily controlled by the permeability contrast between fractures and matrix. Because fracture permeability is governed by fracture aperture, variations in aperture strongly influence the critical head. Equation 17 further indicatesthat fracture retention parameters and inverse air-entry pressure directly affect the critical head, whereas matrix retention parameters do not play a role. Sensitivity analyses confirm this (Text S2 and Figures S1--S7 in Supporting Information). This is expected because the transition occurs after the matrix becomes fully saturated, when its relative permeability reaches unity and matrix retention parameters no longer play a role. It should be noted that this analytical prediction assumes well-connected fracture networks; in disconnected networks, matrix properties may also influence the transition.

To assess the generality of our framework, we have additionally tested the van Genuchten retention model (see Text S5 and Figure S12 in Supporting Information). A similar two-branch behavior is observed, with consistent values of $S_c$, $k_{rc}$ and $p_{cc}$ obtained, demonstrating that the framework is general and not restricted to the Brooks--Corey formulation.

\subsection{Control of fracture network connectivity}

The two-branch retention behavior persists across fracture networks with varying connectivity. Re-examination of Equation 17 shows that the critical transition is controlled primarily by the fracture intensity. The percolation parameter quantifying fracture network connectivity does not explicitly appear in the formulation, but it influences the validity of the analytical prediction. As shown in Figure 4, the analytical solution systematically underestimates the dimensionless critical pressure head for disconnected networks. This discrepancy arises from the assumption in Equation 17 that fractures form fully connected pathways for flow. While this assumption may hold for well-connected networks, where multiple flow paths distribute water flux, it breaks down near or below the percolation threshold (Hyman et al., 2019; Davy et al., 2024). In disconnected networks, fractures alone cannot transmit flow through the system, and a substantial fraction of the flux must pass through matrix, thereby involving fracture--matrix exchange (Yoon et al., 2023), consistent with the limited increase in relative permeability between the critical saturation and full saturation (Text S2 and Figure S8 in Supporting Information). The fracture intensity used in Equation 17 therefore overestimates the hydraulically active fracture population in disconnected networks, leading to an underestimation of the dimensionless critical pressure head.

\section{Conclusions}

This study resolves the longstanding debate over the relative importance of matrix and fracture flow in unsaturated fractured rock. Using three-dimensional numerical simulations that explicitly represent interacting fracture--matrix systems, we show that unsaturated flow exhibits an emergent two-branch retention behavior reflecting a transition between matrix- and fracture-dominated regimes. A generalized analytical retention formulation is proposed, capturing the emergence of the two regimes and identifying a critical saturation that delineates them. The regime transition is revealed to be governed primarily by fracture properties, including aperture, air-entry pressure, and fracture intensity, whereas matrix constitutive parameters play a limited role once the matrix becomes fully saturated, particularly in well-connected networks. The two-branch retention behavior persists across fracture networks with varying connectivity, indicating that the regime transition arises fundamentally from fracture--matrix flow partitioning. These results provide a physically grounded basis for upscaling unsaturated flow in fractured rock by capturing the transition between matrix- and fracture-dominated regimes.

\section*{Acknowledgments}

We are grateful for the financial support from the Mont Terri Consortium and Uppsala University. We also thank the Mont Terri PF-A experiment partners Swiss Federal Office of Topography (swisstopo), Swiss Federal Nuclear Safety Inspectorate (ENSI), German Federal Institute for Geosciences and Natural Resources (BGR), and Swiss Seismological Service (SED) for their support. Q.L. is grateful for support from the European Research Council (ERC) under the European Union’s Horizon Europe programme (ERC Consolidator Grant, grant no. 101232311) for the project ``Unified framework for modelling progressive to catastrophic failure in fractured media (FORECAST)''.

\section*{Open Research}

The datasets generated in this study, together with the scripts used for data analysis and figure production, are available in Andiva et al. (2026).

\section*{Conflict of Interest Disclosure}

The authors declare there are no conflicts of interest for this manuscript.

\section*{References}

\begin{list}{}{\setlength{\leftmargin}{1.5em}%
\setlength{\itemindent}{-1.5em}%
\setlength{\itemsep}{0.5em}%
\setlength{\parsep}{0pt}%
\setlength{\topsep}{0pt}}

\item Ababou, R. (2018). Capillary flows in heterogeneous and random porous media. John Wiley \& Sons.

\item Andiva, M. R., Jiang, C., Ziegler, M., \& Lei, Q. (2026). Dataset and code for “Two-branch retention behavior in unsaturated fractured rock driven by fracture-matrix flow partitioning” [Dataset]. Zenodo. \url{https://doi.org/10.5281/ZENODO.19346310}

\item Assouline, S. (2005). On the relationships between the pore size distribution index and characteristics of the soil hydraulic functions. Water Resources Research, 41(7). \url{https://doi.org/10.1029/2004WR003511}

\item Balberg, I. (1985). Universal percolation-threshold limits in the continuum. Physical Review B, 31(6), 4053--4055. \url{https://doi.org/10.1103/PhysRevB.31.4053}

\item Bear, J. (1972). Dynamics of fluids in porous media. Elsevier.

\item Birkholzer, J., Li, G., Tsang, C.-F., \& Tsang, Y. (1999). Modeling studies and analysis of seepage into drifts at Yucca mountain. Journal of Contaminant Hydrology, 38(1--3), 349--384. \url{https://doi.org/10.1016/S0169-7722(99)00020-0}

\item Bodvarsson, G. S., Wu, Y.-S., \& Zhang, K. (2003). Development of discrete flow paths in unsaturated fractures at Yucca Mountain. Journal of Contaminant Hydrology, 62--63, 23--42. \url{https://doi.org/10.1016/S0169-7722(02)00177-8}

\item Bonnet, E., Bour, O., Odling, N. E., Davy, P., Main, I., Cowie, P., \& Berkowitz, B. (2001). Scaling of fracture systems in geological media. Reviews of Geophysics, 39(3), 347--383. \url{https://doi.org/10.1029/1999RG000074}

\item Bour, O., \& Davy, P. (1998). On the connectivity of three-dimensional fault networks. Water Resources Research, 34(10), 2611--2622. \url{https://doi.org/10.1029/98WR01861}

\item Broadbridge, P., \& White, I. (1988). Constant rate rainfall infiltration: A versatile nonlinear model, 1. Analytical solution. Water Resources Research, 24, 145--154.

\item Brooks, R. H., \& Corey, A. T. (1966). Properties of porous media affecting fluid flow. J. Irrig. Drain. Div., 92, 61--88.

\item Cey, E., Rudolph, D., \& Therrien, R. (2006). Simulation of groundwater recharge dynamics in partially saturated fractured soils incorporating spatially variable fracture apertures. Water Resources Research, 42(9). \url{https://doi.org/10.1029/2005WR004589}

\item Davy, P., Le Goc, R., Darcel, C., Pinier, B., Selroos, J.-O., \& Le Borgne, T. (2024). Structural and hydrodynamic controls on fluid travel time distributions across fracture networks. Proceedings of the National Academy of Sciences, 121(47), e2414901121. \url{https://doi.org/10.1073/pnas.2414901121}

\item de Dreuzy, J.-R., Davy, P., \& Bour, O. (2000). Percolation parameter and percolation-threshold estimates for three-dimensional random ellipses with widely scattered distributions of eccentricity and size. Physical Review E, 62(5), 5948--5952. \url{https://doi.org/10.1103/PhysRevE.62.5948}

\item Demond, A. H., \& Roberts, P. V. (1991). Effect of interfacial forces on two-phase capillary pressure—Saturation relationships. Water Resources Research, 27(3), 423--437. \url{https://doi.org/10.1029/90WR02408}

\item Gershenzon, N. I., Ritzi Jr., R. W., Dominic, D. F., Mehnert, E., \& Okwen, R. T. (2016). Comparison of CO2 trapping in highly heterogeneous reservoirs with Brooks-Corey and van Genuchten type capillary pressure curves. Advances in Water Resources, 96, 225--236. \url{https://doi.org/10.1016/j.advwatres.2016.07.022}

\item Glass, R. J., Nicholl, M. J., Rajaram, H., \& Wood, T. R. (2003). Unsaturated flow through fracture networks: Evolution of liquid phase structure, dynamics, and the critical importance of fracture intersections. Water Resources Research, 39(12). \url{https://doi.org/10.1029/2003WR002015}

\item Glass, R. J., Nicholl, M. J., \& Tidwell, V. C. (1995). Challenging models for flow in unsaturated, fractured rock through exploration of small scale processes. Geophysical Research Letters, 22(11), 1457--1460. \url{https://doi.org/10.1029/95GL01490}

\item Huang, H., Meakin, P., Liu, M., \& McCreery, G. E. (2005). Modeling of multiphase fluid motion in fracture intersections and fracture networks. Geophysical Research Letters, 32(19). \url{https://doi.org/10.1029/2005GL023899}

\item Hyman, J. D., Dentz, M., Hagberg, A., \& Kang, P. K. (2019). Linking structural and transport properties in three-dimensional fracture networks. Journal of Geophysical Research: Solid Earth, 124(2), 1185–1204. \url{https://doi.org/10.1029/2018JB016553}

\item Ji, S.-H., Nicholl, M. J., Glass, R. J., \& Lee, K.-K. (2004). Influence of a simple fracture intersection on density-driven immiscible flow: Wetting vs. nonwetting flows. Geophysical Research Letters, 31(14). \url{https://doi.org/10.1029/2004GL020045}

\item LaViolette, R. A., Glass, R. J., Wood, T. R., McJunkin, T. R., Noah, K. S., Podgorney, R. K., Starr, R. C., \& Stoner, D. L. (2003). Convergent flow observed in a laboratory-scale unsaturated fracture system. Geophysical Research Letters, 30(2). \url{https://doi.org/10.1029/2002GL015775}

\item Li, Y., Li, X., Teng, S., \& Xu, D. (2014). Improved models to predict gas–water relative permeability in fractures and porous media. Journal of Natural Gas Science and Engineering, 19, 190--201. \url{https://doi.org/10.1016/j.jngse.2014.05.006}

\item Liang, X., Wang, C.-Y., Ma, E., \& Zhang, Y.-K. (2022). Effects of unsaturated flow on hydraulic head response to earth tides–an analytical model. Water Resources Research, 58(2), e2021WR030337. \url{https://doi.org/10.1029/2021WR030337}

\item Liu, H. H., Bodvarsson, G. S., \& Finsterle, S. (2002). A note on unsaturated flow in two-dimensional fracture networks. Water Resources Research, 38(9), 15-1-15--19. \url{https://doi.org/10.1029/2001WR000977}

\item Liu, H. H., Doughty, C., \& Bodvarsson, G. S. (1998). An active fracture model for unsaturated flow and transport in fractured rocks. Water Resources Research, 34(10), 2633--2646. \url{https://doi.org/10.1029/98WR02040}

\item Liu, H.-H., \& Bodvarsson, G. S. (2001). Constitutive relations for unsaturated flow in a fracture network. Journal of Hydrology, 252(1), 116--125. \url{https://doi.org/10.1016/S0022-1694(01)00449-8}

\item Nicholl, M. J., Glass, R. J., \& Wheatcraft, S. W. (1994). Gravity-driven infiltration instability in initially dry nonhorizontal fractures. Water Resources Research, 30(9), 2533--2546. \url{https://doi.org/10.1029/94WR00164}

\item Nimmo, J. R. (2012). Preferential flow occurs in unsaturated conditions. Hydrological Processes, 26(5), Article 5. \url{https://doi.org/10.1002/hyp.8380}

\item Nitao, J. J., \& Buscheck, T. A. (1991). Infiltration of a liquid front in an unsaturated, fractured porous medium. Water Resources Research, 27(8), 2099--2112. \url{https://doi.org/10.1029/91WR01369}

\item Peters, R. R., \& Klavetter, E. A. (1988). A continuum model for water movement in an unsaturated fractured rock mass. Water Resources Research, 24(3), 416--430. \url{https://doi.org/10.1029/WR024i003p00416}

\item Pruess, K. (1998). On water seepage and fast preferential flow in heterogeneous, unsaturated rock fractures. Journal of Contaminant Hydrology, 30(3--4), 333--362. \url{https://doi.org/10.1016/S0169-7722(97)00049-1}

\item Pruess, K., Faybishenko, B., \& Bodvarsson, G. S. (1999). Alternative concepts and approaches for modeling flow and transport in thick unsaturated zones of fractured rocks. Journal of Contaminant Hydrology, 38(1), Article 1. \url{https://doi.org/10.1016/S0169-7722(99)00018-2}

\item Pruess, K., \& Tsang, Y. W. (1990). On two‐phase relative permeability and capillary pressure of rough‐walled rock fractures. Water Resources Research, 26(9), 1915--1926. \url{https://doi.org/10.1029/WR026i009p01915}

\item Rogers, C., Stallybrass, M. P., \& Clements, D. L. (1983). On two phase filtration under gravity and with boundary infiltration: Application of a Backlund transformation. Nonlinear Analysis, Theory, Methods and Applications, 7, 785--799.

\item Rüdiger, F., Dentz, M., \& Kordilla, J. (2024). Partially saturated fracture-matrix infiltration in experiments and theory. Water Resources Research, 60(7), e2023WR036323. \url{https://doi.org/10.1029/2023WR036323}

\item Rutqvist, J., \& Tsang, C.-F. (2012). Multiphysics processes in partially saturated fractured rock: Experiments and models from Yucca Mountain. Reviews of Geophysics, 50(3). \url{https://doi.org/10.1029/2012RG000391}

\item Seol, Y., Liu, H. H., \& Bodvarsson, G. S. (2003). Effects of dry fractures on matrix diffusion in unsaturated fractured rocks. Geophysical Research Letters, 30(2). \url{https://doi.org/10.1029/2002GL016118}

\item Sherman, T., Hyman, J., Dentz, M., \& Bolster, D. (2020). Characterizing the influence of fracture density on network scale transport. Journal of Geophysical Research: Solid Earth, 125(1), e2019JB018547. \url{https://doi.org/10.1029/2019JB018547}

\item Tidwell, V. C., Glass, R. J., \& Peplinski, W. (1995). Laboratory investigation of matrix imbibition from a flowing fracture. Geophysical Research Letters, 22(11), 1405--1408. \url{https://doi.org/10.1029/95GL01097}

\item Tokunaga, T. K., Olson, K. R., \& Wan, J. (2005). Infiltration flux distributions in unsaturated rock deposits and their potential implications for fractured rock formations. Geophysical Research Letters, 32(5). \url{https://doi.org/10.1029/2004GL022203}

\item van Genuchten, M. Th. (1980). A closed for equation for predicting the hydraulic conductivity of unsaturated soils. Soil Sci. Soc., 44, 892--898.

\item Viswanathan, H. S., Ajo-Franklin, J., Birkholzer, J. T., Carey, J. W., Guglielmi, Y., Hyman, J. D., Karra, S., Pyrak-Nolte, L. J., Rajaram, H., Srinivasan, G., \& Tartakovsky, D. M. (2022). From fluid flow to coupled processes in fractured rock: recent advances and new frontiers. Reviews of Geophysics, 60(1), e2021RG000744. \url{https://doi.org/10.1029/2021RG000744}

\item Wang, C.-Y., Zhu, A.-Y., Liao, X., Manga, M., \& Wang, L.-P. (2019). Capillary effects on groundwater response to earth tides. Water Resources Research, 55(8), 6886--6895. \url{https://doi.org/10.1029/2019WR025166}

\item Wang, J. S. Y., \& Narasimhan, T. N. (1985). Hydrologic mechanisms governing fluid flow in a partially saturated, fractured, porous medium. Water Resources Research, 21(12), 1861--1874. \url{https://doi.org/10.1029/WR021i012p01861}

\item Witherspoon, P. A., Wang, J. S. Y., Iwai, K., \& Gale, J. E. (1980). Validity of Cubic Law for fluid flow in a deformable rock fracture. Water Resources Research, 16(6), 1016--1024. \url{https://doi.org/10.1029/WR016i006p01016}

\item Wood, T. R., Nicholl, M. J., \& Glass, R. J. (2002). Fracture intersections as integrators for unsaturated flow. Geophysical Research Letters, 29(24), 44-1-44--4. \url{https://doi.org/10.1029/2002GL015551}

\item Xiang, L., Li, C., \& Meng, J. (2025). Effect of matrix permeability on non-Darcian flow behavior and flow partitioning patterns in fractured porous media: Insight from experimental and numerical tests. Journal of Hydrology, 658, 133234. \url{https://doi.org/10.1016/j.jhydrol.2025.133234}

\item Xu, B., Yortsos, Y. C., \& Salin, D. (1998). Invasion percolation with viscous forces. Physical Review E, 57(1), 739--751. \url{https://doi.org/10.1103/PhysRevE.57.739}

\item Yoon, S., Hyman, J. D., Han, W. S., \& Kang, P. K. (2023). Effects of dead-end fractures on non-fickian transport in three-dimensional discrete fracture networks. Journal of Geophysical Research: Solid Earth, 128(7), e2023JB026648. \url{https://doi.org/10.1029/2023JB026648}

\item Zhang, C.-L. (2018). Thermo-hydro-mechanical behavior of clay rock for deep geological disposal of high-level radioactive waste. Journal of Rock Mechanics and Geotechnical Engineering, 10(5), 992--1008. \url{https://doi.org/10.1016/j.jrmge.2018.03.006}

\item Zhou, Z., Yang, Z., Xue, S., Hu, R., \& Chen, Y.-F. (2022). Liquid breakthrough time in an unsaturated fracture network. Water Resources Research, 58(3), e2021WR031012. \url{https://doi.org/10.1029/2021WR031012}

\item Zhu, W., Khirevich, S., \& Patzek, T. W. (2021). Impact of fracture geometry and topology on the connectivity and flow properties of stochastic fracture networks. Water Resources Research, 57(7), e2020WR028652. \url{https://doi.org/10.1029/2020WR028652}

\end{list}

\end{document}